\documentclass{article}
\usepackage[utf8]{inputenc}
\usepackage[]{cite}
\usepackage{graphicx}
\usepackage{authblk}
\usepackage{hyperref}
\usepackage{amsmath}
\providecommand{\keywords}[1]{\textbf{\textit{Keywords --}} #1}
\begin{document}

\title{Testing the Dark Universe through the Layzer-Irvine Equation}

\author[1,2]{C. Gomes \footnote{\href{mailto:claudio.gomes@fc.up.pt}{claudio.gomes@fc.up.pt}}}

\affil[1]{Centro de Física das Universidades do Minho e do Porto, Faculdade de Ciências da Universidade do Porto, Rua do Campo Alegre s/n, 4169-007 Porto
}
\affil[2]{Institute Okeanos - University of the Azores, Campus da Horta, Rua Professor Doutor Frederico Machado 4, 9900-140 Horta, Portugal\\
}

\maketitle

\begin{abstract}
We review the cosmic generalisation of the virial theorem known as the Layzer-Irvine equation, also independently derived by Dmitriev and Zeldovich. This equation has been studied in the literature for dark matter-dark energy interaction models, as well as in the context of alternative theories of gravity. We discuss results from the previous scenarios and point out future directions.
\end{abstract}

\keywords{Virial Theorem; Dark Matter; Dark Energy; Modified Gravity.}

\section{Introduction}

More than a century from being proposed, Einstein's General Theory of Relativity (GR) stands as a central piece in describing gravity, meeting available data with incredible precision \cite{CMWill}. However, at large scales it requires two major presently unknown energy components to match data, namely dark matter and dark energy. Moreover, at small scales, it is not fully compatible with Quantum Mechanics. These issues, in addition to the well-known cosmological constant problem, and the presence of singularities, led to alternative gravity scenarios \cite{Capozziello:2011,Odintsov:2011,Odintsov:2017}.

In particular, GR may be extended to an arbitrary function of the scalar curvature in the so-called $f(R)$ theories \cite{fR1,fR2}. This can be further generalised to account for an extra function of the curvature nonminimally coupled to the matter Lagrangian density, which is known as non-minimal matter-curvature coupling gravity \cite{nmc}. Other scenarios also exist, where one introduces torsion and/or non-metricity \cite{Trinity,Gomes:2019}, having a metric-Palatini approach \cite{HMP,GHMP}, or having the most general scalar field theory with equations of motion up to second order \cite{Horndeski,DHOST}, to name a few.

In the opposite direction, one may consider that dark matter and dark energy do exist in the form of exotic particles beyond the Standard Model of Particle Physics, or as exotic fluids such as the generalised Chaplygin gas \cite{Bento:2002}. Moreover, the two dark components could interact with each other or arise from a more fundamental unifying framework, even accounting for inflation \cite{Bento:2002,Liddle:2006,Liddle:2008,Henriques:2009,Sa:2020,Aljaf:2020eqh}.

One tool for studying gravitationally bound systems is the virial theorem. In fact, from this Zwicky derived the amount of nonluminous matter that should exist in the Coma Cluster, coining it as Dark Matter \cite{Zwicky}. The virial theorem can be derived in such a way as to have a tensor formulation \cite{BinneyTremaine}. It can also be derived in a cosmological context, namely from cosmological perturbations to account for an expanding Universe, which is known as the Layzer-Irvine equation \cite{Layzer,Irvine,Irvine:1965}, or also Dmitriev-Zeldovich \cite{Zeldovich:1964en}, or the cosmic energy equation. This equation for sufficiently relaxed systems renders the usual virial theorem. Therefore, this framework has been generalised to accommodate dark matter-dark energy interactions \cite{Bertolami:2007, He:2010, Grigoris:2021}, inhomogeneous dark energy \cite{Gomes:2013}, and modified gravity models \cite{Gomes:2014,Winther:2013,Shtanov:2010}, for instance, leading to modified virial equations and hence ways to test these properties in the "missing mass" feature at galaxies and clusters. The virial theorem has also been studied from the Boltzmann equation in the scalar-tensor representation of $f(R)$ theories \cite{VirialFR}. At the same time, the Layzer-Irvine equation has been proven to be fundamental in assessing the accuracy of cosmological N-body simulations in the non-linear regime \cite{Joyce,Benhaiem:2013xfy,List:2023jxz}.

Therefore, the aim of this work is to critically review the Layzer-Irvine equation, discussing the most suitable derivation method depending on the model, and to stress the need for testing galaxy clusters such as Abell 586 cluster using realistic density profiles.

This paper is organised as follows. In Sec. \ref{sec:LIeq} we derive the Layzer-Irvine equation resorting to scalling arguments and to cosmological perturbation theory. In Sec. \ref{sec:GLI}, we analyse generalisations of this equation in the context of dark matter-dark energy models and in modified gravity scenarios. Then, in Sec. \ref{sec:Abell586}, we compare different density profiles with data from the spherically symmetric Abell 586 cluster as tests of the deviation from the standard virial ratio. Finally, we draw our conclusions in Sec. \ref{sec:conclusions}.

\section{The Layzer-Irvine Equation}\label{sec:LIeq}

An intuitive demonstration of the Layzer-Irvine equation starts by considering a local inhomogeneity associated to a system of $N$ interacting particles (as Cold Dark Matter particles) with masses $m_i$, with comoving positions $\Vec{x}_i$ and physical distances $\Vec{r}_i=a(t)\Vec{x}_i$ where $a(t)$ is the scale factor, and momentum $\Vec{p}_i=m_i \Vec{v}_i$, where the peculiar velocity reads $\Vec{v}_i=a(t)\dot{\Vec{x}}_i=\dot{\Vec{r}}_i-H\Vec{r}_i$, where $H=\dot{a}/a$ is the Hubble parameter, and $i=1,...,N$ \cite{Gomes:2013}. Thus, the Hamiltonian of this system is simply:
\begin{equation}
    \mathcal{H}=\mathcal{K}+\mathcal{U}~,
\end{equation}
where the peculiar kinetic energy, $\mathcal{K}$, and the peculiar potential energy, $\mathcal{U}$, are given by:
\begin{eqnarray}
    & \mathcal{K} = \sum_{i=1}^N\frac{p_i^2}{2m_i}~,\\
    & \mathcal{U} = -\frac{G}{2}\int \frac{(\rho(\vec{r})-\bar{\rho})(\rho(\vec{r'})-\bar{\rho})}{\|\vec{r}-\vec{r'}\|}d\vec{r}d\vec{r'}~.
\end{eqnarray}

The classical energy equation is simply $\dot{\mathcal{E}}=\partial\mathcal{E}/\partial t$, it is straightforward to verify that the peculiar kinetic and potential energies scale as $\mathcal{K}\propto a^{-2}$ and $\mathcal{U}\propto a^{-1}$. Therefore, we can write:
\begin{equation}
    \dot{\mathcal{E}}+H\left(2\mathcal{K}+\mathcal{U}\right)=0~.
\end{equation}

This derivation originates from the scaling property of the cosmic equation and of the virial theorem. Another derivation method resorts to cosmological perturbation theory. For this, we consider a homogeneous and isotropic Universe which is well described by a Robertson-Walker metric, $\gamma_{ij}$:
\begin{equation}
    ds^2=-(1+2\Phi)dt^2+(1-2\Psi)\gamma_{ij}dx^idx^j~,
\end{equation}
and the matter-energy content of the Universe is a perfect fluid, $T^{\mu\nu}=\rho ~u^{\mu}u^{\nu}+p ~g^{\mu\nu}$, where we used the 4-velocity by $u^{\mu}=(1,u^i)$, under the normalisation condition $u^{\mu}u_{\mu}=-1$. Perturbing the $0i$-components of the metric field equations, in a matter-dominated and pressure-less Universe, $\rho\approx\rho_m$ and $p\approx 0$, and assuming that $u_i \approx a \dot{x}_i=v_{m~ i}$, we find:
\begin{equation}
    \frac{\partial}{\partial t}(av_m)=-a\nabla_r \Phi~.\label{eqn:0i}
\end{equation}

From the perturbed Einstein's equations, in the subhorizon approximation, $k/a > H$, where $k$ is the wavenumber in Fourier space, we find:
\begin{equation}
    \dot{\sigma}_m+3H\sigma_m=-\frac{1}{a}\nabla_x\cdot\left(\rho_m \vec{v}_m\right)~, \label{eqn:matterperturbations}
\end{equation}
where $\sigma_m=\delta\rho_m$ is the energy density perturbation. Moreover, the background behaves as follows:
\begin{equation}
    \dot{\rho}_m+3H\rho_m=0~.\label{eqn:background}
\end{equation}

Afterwards, we contract Eq. (\ref{eqn:0i}) with $a\vec{v}_m\rho_m$, and integrate over the volume $d^3r$:
\begin{equation}
    \int a\vec{v}_m\rho_m \frac{\partial}{\partial t}(av_m) d^3r=- \int a\vec{v}_m\rho_m \frac{1}{a}\nabla_x\cdot\left(\rho_m \vec{v}_m\right) d^3r~.
\end{equation}

From Eqs. (\ref{eqn:matterperturbations}) and (\ref{eqn:background}), noting that the kinetic energy is $K=1/2\int \rho v_m ^2 d^3r$, and performing integration by parts, one gets:
\begin{equation}
    \frac{\partial K}{\partial t}+2HK=- \int \Phi \left(\sigma_m d^3r\right) ~.
\end{equation}

Since the Bardeen potential obeys the Poisson equation, $\Phi=-G\int \frac{\sigma(r',t)}{\|r-r'\|}d^3r'$, and noting its associated potential energy as $U=1/2\int \Phi \sigma_m d^3r$, after some algebra we get the Layzer-Irvine equation \cite{Layzer,Irvine,Irvine:1965}:
\begin{equation}
    \frac{\partial}{\partial t}(K+U)=-H(2K+U)~.
\end{equation}

\section{Generalising the Layzer-Irvine Equation}\label{sec:GLI}

The cosmic energy equation has been generalised in several contexts, namely in dark matter-dark energy interaction scenarios and in alternative models of gravity. We examine these extensions and their physical implications. 

\subsection{Dark Matter-Dark Energy Models}\label{subsec:DMDE}

Dark matter and dark energy may interact with each other, hence one's dynamics influencing the other. Thus, in an expanding Universe dominated by the two dark components, the Friedmann equations are the same, but the continuity equation is modified \cite{Grigoris:2021,Zhao:2022ycr}:
\begin{eqnarray}
    &&\dot{\rho}_{DM}+3H\rho_{DM}=Q\\
    &&\dot{\rho}_{DE}+3H\rho_{DE}(1+w_{DE})=-Q~,
\end{eqnarray}
where the equation of state parameter for dark matter is zero as it is pressure-less, and $Q$ is the source term responsible for the energy flow between the dark components. The specific form of the source term will depend on the interaction \cite{Bertolami:2007,Abdalla:2007,Abdalla:2009,Aljaf:2020eqh,Zhao:2022ycr}.

\subsubsection{Quintessence model with DE-DM interaction}

In the coupled quintessence model, $Q=\xi H \rho_{DM}$. We can further assume that there is a scaling relation between dark matter and dark energy of the form:
\begin{equation}
    \frac{\rho_{DE}}{\rho_{DM}}=a^\eta\frac{\Omega_{DE}}{\Omega_{DM}}~,
\end{equation}
where $\Omega_i=\frac{\rho_i}{\rho_c}$, $i\in \{DM,DE\}$, with the critical density $\rho_c(t)=\frac{3H^2(t)}{8\pi G}$, often computed at present day with a subscript $0$. Both the exponent $\eta$ and the model parameter $\xi$ are related as \cite{Bertolami:2007}:
\begin{equation}
    \xi=\frac{\eta+3w_{DE}}{1+(\Omega_{DM0}/\Omega_{DM0})a^{-\eta}} ~.
\end{equation}

From scaling arguments, one may straightforwardly check that $\rho_K \sim a^{-2}$, whilst $\rho_U\sim a^{\eta-1}$, hence the Layzer-Irvine equation becomes \cite{Bertolami:2007}:
\begin{equation}
    \dot{\rho}_{DM}+H\left(2\rho_K+(1-\xi)\rho_U\right)=0~,
\end{equation}
from which a relaxed physical system has a virial ratio of:
\begin{equation}
    \frac{\rho_K}{\rho_U}=\frac{\xi-1}{2}~.
\end{equation}

\subsubsection{Generalised Chaplygin gas}

The Generalised Chaplygin gas is a model that attracted a lot of attention in the scientific literature, and assumes an exotic fluid with equation of state:
\begin{equation}
    p=-\frac{A}{\rho^{\alpha}}~,
\end{equation}
where $A>0$ is a constant and $\alpha$ is a real number. The main features were that this fluid could unify the description of the evolution of a cosmological fluid that behaves as dust, matter, or a cosmological constant for different values of $\alpha$.
This model is related to the coupled quintessence model through the relation $\eta=3(1+\alpha)$ \cite{Bertolami:2007}. Therefore, the Layzer-Irvine equation follows from the previous scenario:
\begin{equation}
    \dot{\rho}_{DM}+H\left(2\rho_K+\left(1-\frac{3(1+\alpha + w_{DE})}{1+(\Omega_{DM0}/\Omega_{DM0})a^{-3(1+\alpha)}} \right)\rho_U\right)=0~.
\end{equation}

\subsubsection{Coupled dark matter-dark energy}

We can consider the case where $Q=\xi H (\rho_{DM}+\rho_{DE})$ \cite{Abdalla:2007}. In this case, the Layzer-Irvine equation becomes:
\begin{equation}
    \frac{\partial}{\partial t}\left[K_{DM}+(1+b_{em})U_{DM}\right]+ H\left[(2-\bar{\zeta}) K_{DM}+(1+b_{em})(1-2\bar{\zeta})U_{DM}\right]=0~,
\end{equation}
where the term $b_{em}$ is assumed to be a constant that relates the local densities of dark energy and of dark matter through $\delta\rho_{DE}=b_{em}\delta\rho_{DM}$, and $\bar{\zeta}=(1+b_{em})\zeta$.

\subsubsection{Dark Energy inhomogeneities}

The sound speed of quintessence Dark Energy models is the speed of light, $c$, hence DE perturbations in gravitational collapse are small, but in general they need not be. Let us consider two homogeneous concentric spherical patches, whose dynamics are described by the scale factors $a_1$, which we shall call background patch, and $a_2$, which we call a perturbed patch \cite{Gomes:2013}. Then, the peculiar velocity of a local inhomogeneity will be $\vec{v}_i(\vec{r_2}_i)=\Delta H a_2 \vec{x}_i$, where $\vec{x}$ are the comoving coordinates, such that the peculiar kinetic energy per unit mass and per $x^5$ (the reason for this term will be evidence in the potential energy), and in units of $g$, is:
\begin{equation}
    K=G\frac{\mathcal{K}}{Mx^5}=\frac{G}{2x^5}\left<v_i^2\right>=\frac{3}{10}Ga^2_2 (\Delta H)^2 x^{-3}~,
\end{equation}
where the mass $M=\frac{4\pi}{3}\rho_1r_1^3=\frac{4\pi}{3}\rho_2r_2^3$. The potential energy can also be computed, and can be split into a sum of three parts, $\mathcal{U}_A,\mathcal{U}_B,\mathcal{U}_C$:
\begin{eqnarray}
    \frac{x^5}{G}U_A:=\mathcal{U}_A&&=-\frac{3}{5}\frac{GM_+^2}{r_2} \\
    \frac{x^5}{G}U_B:=\mathcal{U}_B&&= -\frac{3}{2}\frac{GM_-^2}{r_1}\left(1-\left(\frac{r_2}{r_1}\right)^2\right)\left(\frac{M_+}{M_-}-\left(\frac{r_2}{r_1}\right)^3\right)\\
    \frac{x^5}{G}U_c:=\mathcal{U}_C&&= -\frac{3}{4}\frac{GM^2_-}{r_1}\left(1-\left(\frac{r_2}{r_1}\right)^5\right)~,
\end{eqnarray}
where $M_+=\frac{4\pi}{3}\rho_+r^3_2$, $M_-=\frac{4\pi}{3}\rho_-r_1^3$, with $\rho_+=\rho_2-\rho_1$ and $\rho_-=-\rho_1$. Therefore, one can compute the time derivative of $E=\frac{G}{x^5}\mathcal{E}$, and arranging terms we get the Layzer-Irvine equation \cite{Gomes:2013}:
\begin{equation}
    \dot{E}+H_1\left(2(1+\alpha)K+U\right)=0~,
\end{equation}
where $\alpha:=-\frac{1}{H_1}\frac{\Delta f}{\Delta H}$, with $f_i:=\dot{H}_i+H^2_i+\frac{1}{2}a^{-3}_i$, and $i\in \{1,2\}$. Therefore, the existence of dark energy perturbations leads to a modification of the standard Layzer-Irvine equation (and subsequently the virial theorem) particularly associated with the kinetic term. We may further note by resorting to the acceleration equation for the scale factor (the Raychaudhury equation) that for homogeneous dark energy we have a vanishing $\alpha$, whilst for inhomogeneous dark energy, this parameter is nonzero, despite being less than unity\cite{Gomes:2013}.

\subsection{Modified Gravity Models}\label{subsec:MG}

In modified gravity, in general, the most suitable derivation for the Layzer-Irvine equation is by resorting to the cosmological perturbation theory. The reason is to accommodate the freedom of the model to later constrain it. However, if one starts by a {\it a priori} defined model, then the derivation by scaling arguments can be easier. We shall look into two models of gravity: a scalar-tensor model, and an extension of $f(R)$ theories of gravity with a non-minimal matter-curvature coupling.

\subsubsection{Scalar-Tensor Gravity}

We consider an action of the form \cite{Winther:2013}:
\begin{equation}
    S=\int d^4x \sqrt{-g}\left[\frac{R}{16\pi G}+f(X,\phi)\right]+S_m[A^2(\phi)g_{\mu\nu},\Psi]~,
\end{equation}
where $f$ is a generic function of the scalar field and its kinetic term given by $X=-\frac{1}{2}\partial_{\lambda}\phi\partial^{\lambda}\phi$, and we have rescaled the metric field $g_{\mu\nu}$ by a conformal factor $A(\phi)$, and $\Phi$ are the matter fields.

The coupling of the scalar field to the metric field will induce an extra-force term in the geodesics equation, which in the non-relativistic limit becomes:
\begin{equation}
    \frac{\partial}{\partial t}(a \vec{v})+(a \vec{v})\frac{\partial \log A}{\partial t}=-a \nabla_r (\Phi_N+\log A)~,
\end{equation}
where $\Phi_N$ is the Newtonian potential, and $\log A$ behaves as an extra potential.
Thus, the Poisson equation is modified:
\begin{equation}
    \nabla_r\Phi_N=4\pi G \left(\delta\rho_J+(1+3c_s^2)\delta\rho_{\phi}\right)~,
\end{equation}
where $\delta\rho_J=A(\phi)\rho_m-\bar{A}(\phi)\bar{\rho}_m$ due to the coupling of $\phi$ with matter fields, an over-bar represents a quantity evaluated in the background, $c_s^2=\frac{\partial p_{\phi}}{\partial \rho_{\phi}}$ is the sound speed associated with the scalar field, and $p_{\phi},\rho_{\phi}$ are the pressure and energy density of the scalar field.
From the Bianchi identities, there follows a conservation equation. In this case, the energy-momentum tensor of matter fields is not conserved (although an effective energy-momentum tensor could be defined to be conserved). Perturbing this equation and subtracting the background evolution equation, we get:
\begin{equation}
    \frac{\left(a^3\dot{\delta \rho}\right)}{a^3}+\nabla_r(\rho_J \vec{v})-\rho_J \vec{v}\nabla_r\log A-\dot{(\log A)}\delta\rho_J -\bar{\rho}_J\frac{\partial}{\partial t}\log\frac{A}{\bar{A}}=0~.
\end{equation}

Contracting the geodesic equation in the non-relativistic limit with $a~\vec{v} \rho_J d^3r $, and following the procedure outlined in the second method explained in Sec. \ref{sec:LIeq}, we get \cite{Winther:2013}:
\begin{eqnarray}
 &&\frac{\partial }{\partial t} \left(K+U_N+U_{\log A}+U_A+U_{\nabla\phi}+U_f+U_{\dot{\phi}}\right) + H\left(2K+U-U_{\nabla\phi}-3U_f+3U_{\dot{\phi}}\right)\nonumber\\
 &&+H\left(\delta K-2\delta U_N-\delta U_{log A}-\delta U_A\right)=0~,   
\end{eqnarray}
where the usual kinetic and potential terms are computed considering $\delta\rho_J$ and the Newtonian potential $\Phi_N$, and the extra terms are:
\begin{align}
U_{\log A} &= -\frac{1}{4\pi G}\int d^3r \left(\nabla_r\Phi_N\right)\cdot(\nabla_r\log A)\\
U_A &= \int d^3r \left(\log A(\phi)-\log A(\overline{\phi})\right)\overline{\rho}_J\\
U_{\nabla\phi} &= \int d^3r f_X\frac{1}{2}(\nabla_r\phi)^2\\
U_{f} &= \int d^3r \left(g(X,\phi)-g(\overline{X},\overline{\phi})\right)\\
U_{\dot{\phi}} &= \int d^3r f_X\frac{1}{2}\left(\dot{\phi}^2-\dot{\overline{\phi}}^2\right)\\
\delta K &= \int d^3r \frac{1}{2}v^2\rho_J \left(\frac{\partial \log A}{\partial \log a}\right)\\
\delta U_N &= \frac{1}{4\pi G}\int d^3r \frac{\Phi_N}{2} \nabla_r^2\Phi_N\frac{\partial\log A}{\partial \log a}\\
\delta U_{\log A} &= \frac{1}{4\pi G}\int d^3r\log A \nabla_r^2\Phi_N\frac{\partial\log A}{\partial \log a}\\
\delta U_A &= \int d^3r \left(\log A(\phi)-\log A(\overline{\phi})\right)\overline{\rho}_J\left(\frac{\partial \log \overline{A}}{\partial \log a}\right)
\end{align}
with the $g$ function defined as $g(X,\phi) \equiv  f_X(X,\phi)X -f(X,\phi)$.

We note that this result is quite cumbersome due to the conformal transformation, which induces coupling terms and perturbations at linear order that contribute mostly for the potential energy.

\subsubsection{Non-minimal Matter-Curvature Coupling Model}
The non-minimal matter-curvature coupling gravity model is given by the action:
\begin{equation}
    S=\int d^4 x\sqrt{-g}\left(\frac{1}{2}f_1(R)+(1+f_2(R))\mathcal{L}\right)~,
\end{equation}
where both $f_1,f_2$ are functions of the Ricci scalar. This model renders a non-conservation law for the energy-momentum tensor of matter fields. Thus, from $\delta T^0_i$ perturbations in that equation and in the field equations, we get the following equation:
\begin{equation}
    \frac{\partial}{\partial t}(a v_m)=-a\nabla_r\left(\Phi+\delta\Phi_c-\dot{\Phi}_c v\right)~,
\end{equation}
where $\Phi_c=\ln f_2$, and $v$ is the potential velocity of the 4-velocity $u_i=-\partial_iv$. We note that we have not imposed the non-relativistic limit, hence $\delta \ln f_2$ which plays the role of a cosmological potential, in contrast with the Newtonian limit where the term $\ln f_2$ is of the same order as the Newtonian potential \cite{BertolamiMartins:2011,Ferreira:2019a,Ferreira:2019b,Bertolami:2020boltzmann,Gomes:2020jeans,March:2021,Gomes:2022quantumjeans}. This assumption is most suitable since it aims at dealing with clusters of galaxies, in opposition to the Newtonian limit for simulations up to astrophysical scales. Moreover, at cosmological scales we have a further term associated with the coupling effect, namely $\dot{\Phi}_cv$
We set $f_1=R$ as to examine the effects of a pure non-minimal coupling, and assume subhorizon approximation, $k/a>H$, where $k$ is the wavenumber in Fourier space. Contracting the previous equation with $a\vec{v}_m\rho_md^3r$, integrating over the volume, and by parts, we get:
\begin{equation}
    \frac{\partial}{\partial t}(K+U+U_{NMC})+H(2K+U+U_{NMC})=0~,
\end{equation}
where $U_{NMC}=\frac{1}{2}\int d^3 r\left(\delta\Phi_c-\dot{\Phi}_cv\right)\delta\rho_m$ is the potential energy associated with the non-minimal coupling.

\section{Constraining with galaxy clusters: the case of Abell 586 cluster}\label{sec:Abell586}

The Abell 586 cluster of galaxies is found at the redshift $z=0.17$. It is sufficiently spherically symmetric in X-ray surface brightness distribution and without any offset between its X-ray and optical centroids, and relaxed in the sense that it has not undergone any merging process in the last Gyrs \cite{Cypriano:2005}. Given its properties and precise observed data from velocity dispersions and mass profiles, it has been used for testing gravity, and the dark sector. In particular, it is a remarkable laboratory to test the equivalence principle, where deviations from the standard virial theorem suggest either an interaction between dark components \cite{Bertolami:2007,Bertolami:2007tq,Bertolami:2012yp,Bahrami:2018ngq,Wang:2008aa}, or modified gravity \cite{Gomes:2013,Gomes:2014}.

Observations from Chandra and Gemini missions have found that the Abell 586 cluster has \cite{Cypriano:2005}:
\begin{enumerate}
    \item a total baryonic mass  $M_{bar} = M_{gas} (1+0.16 \sqrt{h_{70}})$, where the intercluster gas mass amounts to $M_{gas} = 0.48 \times 10^{14} M_{\odot}$, with $M_{\odot}$ being the Solar mass, and the reduced Hubble parameter $h_{70}=H_0/70$, with $H_0$ being the Hubble parameter at present time;
    \item a velocity dispersion with values obtained from distinct methods as in Table \ref{tab:velocities};
    \item a radius of $R=422~kpc$.
\end{enumerate}

\begin{table}[h!]
\begin{tabular}{ |p{4cm}|p{3cm}|  }
\hline
Method & $\sigma_v (km/s)$  \\
\hline
x-ray luminosity\hphantom{00} & $1015 \pm 500$ \\
x-ray temperature\hphantom{00} & $1174 \pm 130$ \\
weak lensing\hphantom{0} & $1243 \pm 58$  \\
velocity distribution & $161 \pm 196$ \\
\hline
\end{tabular}
\caption{Velocity dispersion data from different observation methods of Abell 586 \cite{Cypriano:2005}.}
\label{tab:velocities}
\end{table}

To apply the modified virial theorem to this cluster, one must first find the most suitable density profiles. We shall consider three profiles: top hat, Navarro-Frenk-White (NFW), and isothermal spheres. We note that the NFW profile is somewhat inaccurate for galaxies clusters as it was derived taking into account numerical simulations for galaxies. However, we shall use it to validate the analysis with the other two density profiles.

The top hat density profile corresponds to a constant profile over the radius, whose kinetic over potential energies ratio is given by:
\begin{equation}
    \frac{K}{U}=\frac{\rho_K}{\rho_U}=-3\frac{\sigma_v^2\left<R\right>}{GM}~,
\end{equation}
where $\left<R\right>$ is the mean intergalactic radius.

In its turn, the Navarro-Frenk-White density profile \cite{NFW} leads to a ratio:
\begin{equation}
    \frac{K}{U} \equiv \frac{\rho_K}{\rho_W} =  -\frac{3}{2} \frac{\sigma_v ^2r_0}{GM} \frac{\left[\left(1+\frac{R}{r_0}\right)\ln\left(1+\frac{R}{r_0}\right)-\frac{R}{r_0}\right]^2}       {\left[\left(1+\frac{R}{r_0}\right)\left[\frac{1}{2}\left(1+\frac{R}{r_0}\right)-\ln\left(1+\frac{R}{r_0}\right)\right]-\frac{1}{2}\right]} ~,
\end{equation}
where $\rho_0$ and $r_0$ are the characteristic density and radius, also known as the density and shape parameters of the model.

Finally, the isothermal spheres profile leads to:
\begin{equation}
    \frac{K}{U} \equiv \frac{\rho_K}{\rho_W} =- \frac{3}{2} \frac{\sigma_v ^2R}{G M} ~.
\end{equation}

For the Abell 586 cluster, the mean intergalactic radius is $\left<R\right>=223.6~kpc$ \cite{Bertolami:2012yp}. We can estimate the virial ratio, $\rho_K/\rho_U$, which can be summarised in Tab. \ref{tab:virialratio}.

\begin{table}[h!]
\begin{tabular}{ |p{3.15cm}|p{2.5cm}|p{2.5cm}|p{2.5cm}|  }
\hline
Method & Top Hat & NFW & Isothermal Spheres \\
\hline
x-ray luminosity & $-0.516 \pm 0.516 $ & $-0.408 \pm 0.407 $ &$-0.353 \pm 0.352$\\
x-ray temperature & $-0.691 \pm 0.190 $ & $-0.545 \pm 0.150$& $-0.472 \pm 0.130$\\
weak lensing & $-0.774 \pm 0.145 $  & $-0.611 \pm 0.115$& $-0.529 \pm 0.099$\\
velocity distribution & $-0.676 \pm 0.253$ & $-0.533 \pm 0.200$& $-0.461 \pm 0.173$\\ 
\hline
\end{tabular}
\caption{Virial ratio for Abell 586 cluster from different density profiles as function of the observation methods \cite{Cypriano:2005,Bertolami:2012yp}.}
\label{tab:virialratio}
\end{table}

The deviations from $-1/2$ on the virial ratio of Tab. \ref{tab:virialratio} may signal the existence of an interaction on the dark sector\cite{Bertolami:2007,Bertolami:2012yp} or the effects of modified gravity \cite{Gomes:2013}. One may be careful when looking into data from the weak lensing method as it is biased in the sense that it has correlations between estimated mass and velocity. 

\section{Conclusions}\label{sec:conclusions}

The Layzer-Irvine equation equation is a natural generalisation of the virial theorem to account for the Universe dynamics at large scales. In particular, cosmological structures are studied, such as galaxy clusters. This equation was also independently derived by Dmitriev and Zel'dovich.

This equation is crucial for cosmological simulations that look into deviations from General Relativity. In particular, for sufficiently relaxed physical systems, the former reduces to a modification of the virial theorem, which is a test of the presence of dark matter as Zwicky pioneered, and of possible dark matter-dark energy interaction, or the existence of theories beyond GR.

Most dark matter-dark energy interaction models have one or two free models which can be successfully constrained by data. Moreover, in general, alternative theories of gravity have more than one unfixed parameter as they are formulated in the most general scenario, which may seem arbitrary; however much effort must be done in carrying out a complete analysis of the subsets compatible with different astrophysical and cosmological data, so that the models can be correctly constrained. Or, at least, consistency tests must be performed.

Many models have been used in the study of the virial deviation in clusters, but many more need to be further scrutinised. In particular, in what concerns modified gravity models, extensions with torsion and non-metricity should be compared to assess the fundamentals of spacetime. In addition, future missions such as SKAO and Euclid, among others, will likely find more nearly spherically symmetric galaxy clusters which are needed to have statistics and build catalogs of scenarios being tested.

\section*{Acknowledgments}

C.G. acknowledges support from Fundação para a Ciência e a Tecnologia through UIDB/04650/2025 - Centro de Física das Universidades do Minho e do Porto, and funding from Project 2024.00252.CERN, and European Commission through COST Action CA21136 – “Addressing observational tensions in cosmology with systematics and fundamental physics (CosmoVerse)”, supported by COST (European Cooperation in Science and Technology).

\end{document}